\documentclass[apj]{emulateapj}
\pdfoutput=1 
\usepackage{apjfonts} 
\usepackage[hang,flushmargin]{footmisc}
\usepackage{amsmath,amstext}
\usepackage[breaklinks,colorlinks,citecolor=blue,linkcolor=magenta, urlcolor=red]{hyperref} 
\usepackage[all]{hypcap} 
\usepackage{afterpage}
\usepackage[export]{adjustbox}
\usepackage{soul}
\newcommand{\HST}{{\sl HST }}

\newcommand{\JWST}{{\sl JWST }}

\slugcomment{Accepted for publication in AJ  June 18, 2018}

\shorttitle{WFC3 Transmission Spectrum of WASP-63b}
\shortauthors{Kilpatrick et al.}

\begin{document}

   \title{Community Targets for JWST's Early Release Science Program: Evaluation of WASP-63\MakeLowercase{b}  }

   \author{Brian M. Kilpatrick\altaffilmark{1,}\altaffilmark{$\dagger$},
   Patricio E. Cubillos\altaffilmark{2}, 
   Kevin B. Stevenson\altaffilmark{3},
   Nikole K. Lewis\altaffilmark{3},
   Hannah R. Wakeford\altaffilmark{4},
   Ryan J. MacDonald\altaffilmark{5},
   Nikku Madhusudhan\altaffilmark{5},
   Jasmina Blecic\altaffilmark{6},
   Giovanni Bruno  \altaffilmark{3},
   Adam Burrows\altaffilmark{7},
   Drake Deming\altaffilmark{8},
   Kevin Heng\altaffilmark{9},
   Michael R. Line\altaffilmark{10},
   Caroline V. Morley\altaffilmark{11,}\altaffilmark{$\star$},
   Vivien Parmentier\altaffilmark{12,}\altaffilmark{$\star$},
   Gregory S. Tucker\altaffilmark{1}, 
   Jeff A. Valenti\altaffilmark{3}, 
   Ingo P. Waldmann\altaffilmark{13}, 
   Jacob L. Bean\altaffilmark{14},
   Charles Beichman\altaffilmark{15}, 
   Jonathan Fraine\altaffilmark{3},
   J. E. Krick\altaffilmark{16},
   Joshua D. Lothringer\altaffilmark{12},
   Avi M. Mandell\altaffilmark{17} 
    \\
   }

\affil{1.  Department of Physics, Box 1843, Brown University, Providence, RI 02904, USA}
\affil{2.  Space Research Institute, Austrian Academy of Sciences, Schmiedlstrasse 6, A-8042 Graz, Austria}
\affil{3.  Space Telescope Science Institute, Baltimore, MD 21218, USA}
\affil{4.  Planetary Systems Lab, NASA Goddard Space Flight Center, Greenbelt, MD 20771, USA}
\affil{5.  Institute of Astronomy, University of Cambridge, Madingley Road, Cambridge, CB3 0HA, UK}
\affil{6.  Department of Physics, New York University Abu Dhabi, P.O. Box 129188 Abu Dhabi, UAE}
\affil{7.  Department of Astronomy, University of Maryland, College Park, MD 20742, USA; ddeming@astro.umd.edu}
\affil{8.  Department of Astronomy, University of Maryland, College Park, MD 20742, USA}
\affil{9.  University of Bern, Center for Space and Habitability, Sidlerstrasse 5, CH-3012, Bern, Switzerland}
\affil{10.  School of Earth \& Space Exploration, Arizona State University, Phoenix, AZ 85282, USA}
\affil{11.  Department of Astronomy, Harvard University, Cambridge, MA 02138, USA}
\affil{12.  Lunar \& Planetary Laboratory, University of Arizona, Tucson, AZ 85721, USA}
\affil{13.  Department of Physics \& Astronomy, University College London, Gower Street, WC1E6BT, UK}
\affil{14.  Department of Astronomy and Astrophysics, University of Chicago, 5640 S Ellis Ave, Chicago, IL 60637, USA}

\affil{15.  NASA Exoplanet Science Institute, California Institute of
Technology, Jet Propulsion Laboratory, Pasadena, CA, USA}
\affil{16.  Spitzer Science Center, Infrared Processing and Analysis Center, California Institute of Technology, Mail Code 220-6,
Pasadena, CA 91125, USA}
\affil{17.  Solar System Exploration Division, NASA Goddard Space Flight Center, Greenbelt, MD 20771, USA}
\affil{$\star$ NASA Sagan Fellow}
\affil{$\dagger$ NASA Earth and Space Science Fellow}

\begin{abstract}

We present observations of WASP-63b by the {\sl Hubble Space Telescope (HST)} as part of ``A Preparatory Program to Identify the Single Best Transiting Exoplanet for \JWST Early Release Science". WASP-63b is  one of the community targets under consideration for
the {\sl James Webb Space Telescope (JWST)} Early Release Science (ERS) program. We present a spectrum derived from a single observation by \HST Wide Field Camera~3 in the near infrared.  We engaged groups across the transiting exoplanet community to participate in the analysis of the data and present results from each.  Extraction of the transmission spectrum by several independent analyses find an H$_2$O absorption feature with varying degrees of significance ranging from 1--3 $\sigma$. The feature, in all cases, is muted in comparison to a clear atmosphere at solar composition.  The reasons for the muting of this feature are ambiguous due to a degeneracy between clouds and composition.  The data does not yield robust detections of any molecular species other than H$_2$O. The group was motivated to perform an additional set of retrieval exercises to investigate an apparent bump in the spectrum at $\sim$ 1.55 $\mu$m.  We explore possible disequilibrium chemistry and find this feature is consistent with super-solar HCN abundance but it is questionable if the required mixing ratio of HCN is chemically and physically plausible. The ultimate goal of this study is to vet WASP-63b as a potential community target to best demonstrate the capabilities and systematics of \JWST instruments for transiting exoplanet science.  In the case of WASP-63b, the presence of a detectable water feature indicates that WASP-63b remains a plausible target for \JWST observations.  \\

\end{abstract}

 \maketitle
%

\section{Introduction}\label{sec: Intro}
The {\sl James Webb Space Telescope} will revolutionize transiting exoplanet atmospheric science due to a combination of its capability for continuous, long duration observations and its larger collecting area, spectral coverage, and resolution compared to existing space-based facilities.  We previously outlined a plan in \cite{Commtargs} to fully demonstrate the capabilities of the \JWST instruments during the Early Release Science (ERS) program allowing the community to plan more efficient and successful transiting exoplanet characterization programs in later cycles.\par
	\cite{Commtargs} identified a set of ``community targets" which meet a certain set of criteria for ecliptic latitude, period,  host star brightness, well constrained orbital parameters, and predicted strength of spectroscopic features.  
  WASP-63b was identified as one of the strongest transmission spectroscopy candidates for \JWST Early Release Science. It is an inflated planet ($1.43$ R$_{\rm J}$) with a low mass of only $0.38$ M$_{\rm J}$ \citep{2012Hellier} resulting in a large atmospheric scale height.  It orbits a bright (11.2 V$_{\rm mag}$) star.  Additionally, WASP-63b occupies an important, underexplored, region of transmission spectroscopy due to its mass. Most exoplanets studied in detail with transmission spectroscopy are either hot Jupiters of mass ($\sim$1--3 M$_{\rm J}$)\citep[e.g.][]{2013Deming, 2014Kreidberg, 2015Kreidberg, 2016Line, 2016Sing} or Super Earth-to-Neptune mass planets ($\sim$0.01--0.1 M$_{\rm J}$) \citep[e.g.][]{2014KreidbergGJ1214, 2014Fraine, 2014NatureKnutson}. 
In order to understand formation and evolution processes, it's important to understand the composition of atmospheres over a full and continuous range of masses \citep[e.g.][]{2016Mordasini}.  \par

WASP-63b will be accessible to \JWST approximately six months after the planned April 2019 start of Cycle 1 and ERS observations making it an ideal candidate should there be any delays in the \JWST timetable. Here, we observe WASP-63b to evaluate its suitability as a prime candidate to test the capabilities of {\sl JWST}.   We can use the strength of the water absorption feature at 1.4 $\mu$m as a way to screen potential targets for the presence of obscuring aerosols and determine the amplitude of predicted spectral features \citep[e.g.][]{2013Deming,2014Kreidberg,2016Sing, 2016StevensonClouds}.  Ideally, a clear atmosphere with large amplitude spectroscopic features will be best suited for benchmarking the instruments  and identifying their systematics.

\section{Observations and Analysis}\label{sec: OandA}
We observed the WASP-63 system using the \HST Wide Field Camera 3 (WFC3) on September 19, 2016.  The observations were taken as part of program GO-14642, 
(PI Stevenson).  The observations were made using the G141 grism in forward/reverse spatial scan mode.  Spatial scanning \citep{spatialrep} involves slewing the telescope in the cross dispersion direction during the exposure.  In forward/reverse mode the telescope is exposing in both directions of the slew thus eliminating time to reset the target at the initial position on the detector between exposures.  Each exposure, utilizing SPARS10, consists of 16 non-destructive reads with a total exposure time of approximately 103s which yielded peak per pixel counts near 32,000 electrons.  The spectrum was read out using the 256 $\times$ 256 subarray with a scan rate of $\sim$ 0.08 arcsec/s (0.62 pixels/s).  This corresponds to a total scan length of $\sim$ 8.76 arcsec which spreads the spectrum in the cross dispersion direction over $\sim$ 70 pixels.  We observed WASP-63 for a total of 8 \HST orbits, yielding a total of 86 time series integrations, to cover the entirety of the relatively long duration of transit ($\sim$ 5 hours).  \par

\begin{figure}[!t]
\centering
\includegraphics[trim=0.04in 0.0in 0.0in 0.0in,clip,width=0.48\textwidth, height=0.28\textheight]{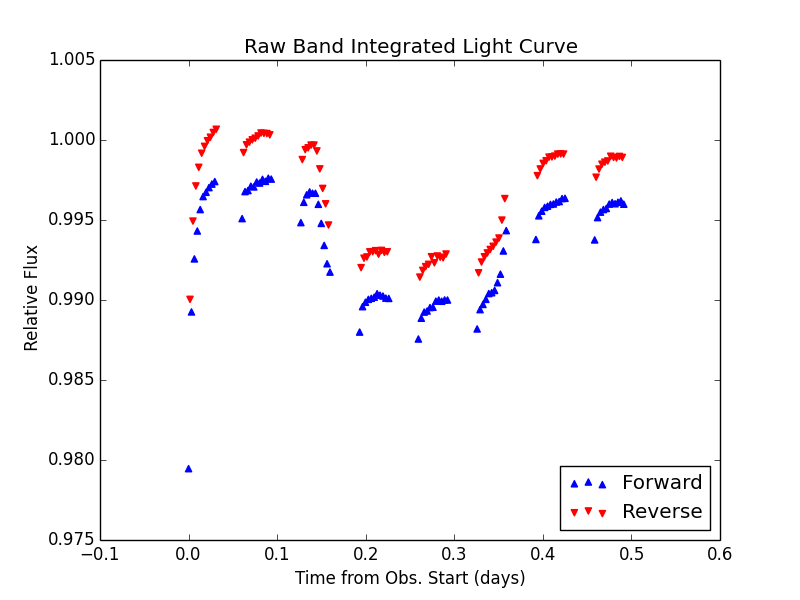}
\includegraphics[trim=0.04in 0.0in 0.0in 0.0in,clip,width=0.48\textwidth, height=0.28\textheight]{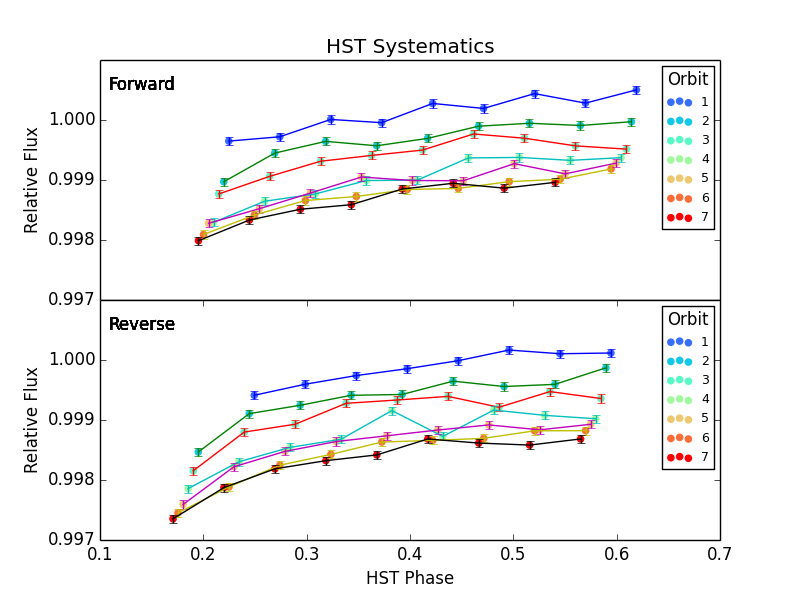} 
\includegraphics[trim=0.04in 0.0in 0.0in 0.0in,clip,width=0.48\textwidth, height=0.29\textheight]{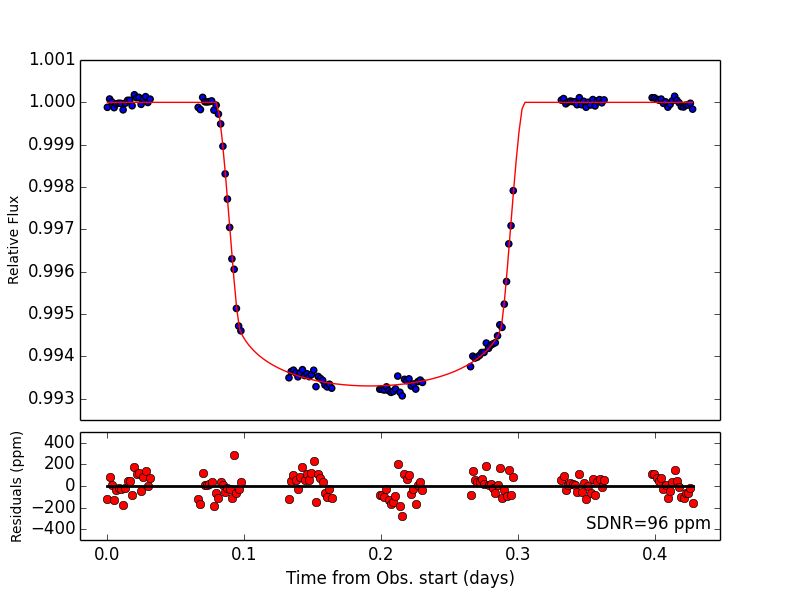}
\caption{{\it Top:}  The normalized raw band integrated light curve.  Forward/Reverse scans are shown in red/blue.  {\it Middle:}  The Band Integrated Light Curve phase folded by \HST Orbital Phase after removing the best fit transit model to illustrate the systematic `hook' in WFC3 observations.  The forward and reverse scan directions are shown in separate panels. Each color corresponds to an \HST orbit. Note that here the first orbit has been discarded since it exhibits different systematics leaving only 7 orbits.  The exponential increase over each orbit and a visit long decrease in response are evident by visual inspection.  {\it Bottom:}  The best fit white light curve shown with systematics removed.  We achieve a standard deviation of the normalized residuals of 96 ppm.}\label{hstsys}
\end{figure}
\afterpage{\clearpage}

  We use the {\sl IMA} files from the CalWF3 pipeline in our analysis.  These files have been calibrated for dark current and zero read bias.  We applied flat field corrections to each non-destructive read (NDR).  Each NDR was background subtracted by considering a background window consisting of ($\sim$50) rows of pixels below the spectrum spanning the dispersion direction.  A mean value for each column of the background window was taken to produce a one-dimensional background correction.  The 1-D solution was then smoothed in the dispersion direction to correct for outliers.  The column by column background value was then subtracted from each pixel of the image.  We then extract the spectrum by taking the difference between successive NDRs.  We apply a top hat filter to each NDR to limit any contribution from cosmic rays and/or overlapping spectra \citep{ 2016Evans, 2017Wakeford}.  The differences between each NDR are then summed to produce a final working image.
  
\subsection{Band Integrated Light Curve}\label{sec: wlc}
We perform the extraction of the band integrated light curve (white light curve) using a range of different aperture sizes in the cross-dispersion direction.  The aperture sizes range from $\pm$10 pixels--$\pm$10 Each orbit includes a direct image of the star from which we calculate the stellar centroid using applying a 2-D Gaussian fit to a 5$\times$5 section of pixels centered on maximum pixel. The trace and wavelength solutions are calculated from the centroid of the undispersed image using the coefficients provided in \cite{2009Kuntschner}.  We extract a box from each working image with the number of columns corresponding to wavelength limits 1.125-1.65 $\mu$m and rows determined by the chosen aperture size.  The band integrated light curve is the summation of all pixels within the box at each time step.  The results of the initial extraction of the raw white light curve are shown in Figure \ref{hstsys} (top). \par
  Fitting the white light curve requires accounting for \HST systematics.  We choose to follow the standard practice of discarding the first orbit as it presents different systematics from the remainder of the data \citep[e.g.][]{2013Deming, 2014Stevenson}.  The raw light curve exhibits a ramp like increase in flux, commonly referred to as the `hook',  with each \HST orbit consistent with previous observations \citep[e.g.][]{2012Berta, 2013Deming,2014Fraine, 2014Kreidberg}.  The hook effect, shown in Figure \ref{hstsys} (middle), is generally steeper in the first frame of each \HST orbit so we discard those data points. We then model the hook as an exponential plus linear function of the form $1-A\exp\left\{S(\theta-\theta_0)\right\}+c_1\theta$ where $\theta$ is the \HST orbital phase, $\theta_0$ is a reference angle for setting zero \HST phase and A, S,  and c$_1$ are scaling factors.  The hook model is combined multiplicatively with a second order polynomial in time over the entirety of the observation.  \par 
  We model the transit using the methods of \cite{MandelAgol} implemented by the Python routine BATMAN \cite{2015BATMAN}.  The transit model assumes a circular orbit with a fixed period. Orbital parameters used for the transit model were taken from \cite{2012Hellier}.  We calculate non--linear limb darkening coefficients  using the PHOENIX Code to fit theoretical spectra as described in detail in \cite{2016Trappist}.  We assume a stellar T$_{eff}$ of 5550$\pm$100 K and log(g) of 4.01 +0.02/-0.04 producing coefficients c$_1$...c$_4$ of (0.36627439,  0.63188403, -0.69135111,  0.23393244) for a limb darkening law of the form $I(\mu)=I_0[1-c_1\left(1-\mu^{1/2}\right)-c_2\left(1-\mu\right)]-c_3\left(1-\mu^{3/2}\right)$ $-c_4\left(1-\mu^{2}\right)$.  During the fitting process we allow for the time of transit center (T$_{\rm cen}$), planetary radius as a fraction of stellar radius (R$_p$/R$_\star$), a/R$_\star$,  inclination (i), and a normalizing factor (the ratio of the amplitudes of the scan directions) to be free parameters and fit both scan directions simultaneously. All fits and uncertainty estimates are derived from the Python routine `emcee' \citep{emcee} utilizing 50 walkers with $10^4$ steps.  Each walker is initialized at a point in parameter space determined by randomly shifting the value of each parameter by some fraction of twice the corresponding one sigma uncertainty.  This forms an over--distributed sphere of starting points surrounding the initial best fit value.   Given the number of observational data during both ingress and egress we place uninformative priors on the orbital properties. The first 20\% of steps are removed for burn in. We test for convergence using Gelman Rubin statistics with a threshold for acceptance of 1.1 and find convergence to levels of 1.01--1.03 in all cases \citep{gelman1992}. We choose the best aperture by minimizing the scatter of the residuals of the white light curve fit. We find a best aperture of $\pm$ 44 pixels in the spatial direction centered on the spectral image centroid.  We achieve a standard deviation of the normalized residuals (SDNR) of 96 parts per million (ppm).  There is no indication that the selection of aperture size has a statistically significant effect on the transit depth fit.  The standard deviation of the transit depth for the five apertures above and below the best aperture is 93 ppm with an average Rp/Rs of 0.077761.  These results all fit well within our best fit value and uncertainty. The best fit white light curve is shown in Figure \ref{hstsys} (bottom) and the values and uncertainties for the planetary physical properties and ephemeris from the best fit white light curve are presented in Table \ref{ephem_fits}.

\capstartfalse
\begin{deluxetable}{cccc}[]
\tablecaption{Best fit values and uncertainties for the planetary physical properties and ephemeris from the band integrated light curve fit. \label{ephem_fits}}
\tablewidth{0.45\textwidth}
\tablehead{
\colhead{R$_p$/R$_\star$} & \colhead{T$_{cen}(MJD)$} & \colhead{a/R$_\star$} & \colhead{i ($^\circ$)}} \label{wlcpars}
0.077762$\pm^{0.000204}_{0.000183}$      & 57650.435$\pm^{6.97\times10^{-5}}_{5.93\times10^{-5}}$      & 6.633$\pm^{0.031}_{0.015}$       & 88.52 $\pm^{0.26}_{0.12}$  
\enddata
\end{deluxetable}
\capstarttrue

\capstartfalse
\def\arraystretch{1.5}
\begin{deluxetable*}{ccc|ccc|ccc}[]
\tablecaption{Transmission spectrum of WASP-63b measured with \HST
WFC3 G141 grism.\label{spectab}}
\tablehead{  \multicolumn{3}{c}{BMK} &   \multicolumn{3}{c}{KBS}   & 
\multicolumn{3}{c}{HRW} \\ 
\colhead{Wavelength($\mu$m)} & \colhead{R$_p$/R$_\star$} & \colhead{Unc (ppm)}&
\colhead{Wavelength($\mu$m)}& \colhead{R$_p$/R$_\star$} & \colhead{Unc (ppm)} &
\colhead{Wavelength($\mu$m)}& \colhead{R$_p$/R$_\star$} & \colhead{Unc (ppm)}} 

\startdata
1.1425     & 0.07791 & 310 &1.1425  & 0.07819&290&1.1352&0.07810&460\\
1.1775     & 0.07732 & 290 &1.1775  & 0.07801&275&1.1677&0.07740&450\\
1.2125     & 0.07717 & 300 &1.2125  &  0.07785&290&1.2002&0.07762&430\\
1.2475     & 0.07753 & 300 &1.2475  &0.07770&270&1.2327&0.07839&420\\
1.2825     & 0.07812 & 280 &1.2825  & 0.07848&260&1.2651&0.07832&420\\
1.3175     & 0.07789 & 270 &1.3175  & 0.07846&285&1.2977&0.07782&420\\
1.3525     & 0.07847 & 290 &1.3525  &0.07853&265&1.3301&0.07832&420\\
1.3875     & 0.07891 & 300 &1.3875  &0.07861&270&1.3626&0.07921&420\\
1.4225     & 0.07832 & 290 &1.4225  & 0.07826&290&1.3951&0.07880&425\\
1.4575     & 0.07839 & 300 &1.4575  & 0.07849&280&1.4275&0.07848&440\\
1.4925     & 0.07773 & 330 &1.4925  &0.07791&290&1.4600&0.07862&445\\
1.5275     & 0.07865 & 330 &1.5275  &0.07831&295&1.4925&0.07811&450\\
1.5625     & 0.07866 & 370 &1.5625  &0.07833&320&1.5250&0.07908&455\\
1.5975     & 0.07816 & 350 &1.5975  &0.07752&310&1.5575&0.07892&485\\
1.6325     & 0.07751 & 360 &1.6325  &0.07718&320&1.5899&0.07821&490\\
&&&&&&1.6224&0.07789&505
\enddata
\end{deluxetable*}
\capstarttrue

\subsection{Spectral Light Curves}\label{sec: slc}
  
  The spectral light curves are extracted using the same aperture we found to minimize the SDNR of the white light curve. The range of wavelengths included in our aperture are divided into 15 bins of width $0.035$ $\mu$m.  The spectrum from each frame is compared to the spectrum of the first frame using cross correlation in Fourier space to check for any shift in the wavelength--pixel solution.  The shift in wavelength solution throughout the observation was on the order of a few tenths of a pixel.  Each column was summed and weighted by the fraction of that pixel in the bin.  \par
  
    \begin{figure}[]\label{slc_fits}
  \centering
  \begin{minipage}[b]{0.19\textwidth}
    \includegraphics[trim=0.0in 0.0in 0.44in 0.0in,clip,width=\textwidth, height=0.58\textheight]{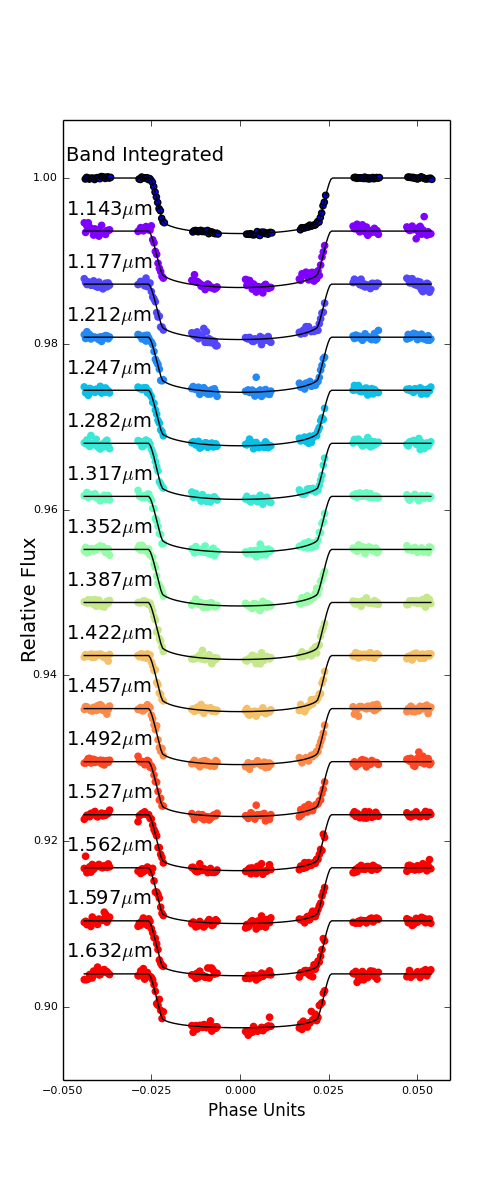}
  \end{minipage}
  \hfill
  \begin{minipage}[b]{0.12\textwidth}
    \includegraphics[trim=0.380in 0.0in 0.25in 0.0in,clip,width=\textwidth, height=0.58\textheight]{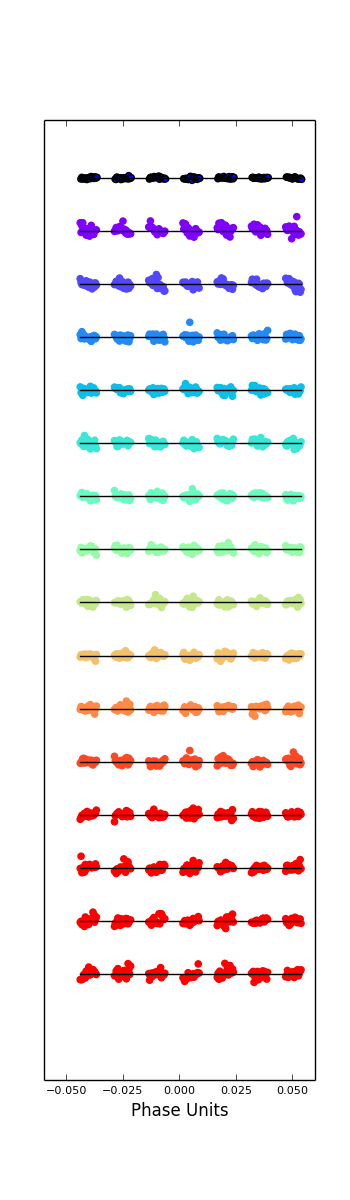}
  \end{minipage}
    \hfill
  \begin{minipage}[b]{0.15\textwidth}
    \includegraphics[trim=0.380in 0.0in 0.05in 0.0in,clip,width=\textwidth, height=0.58\textheight]{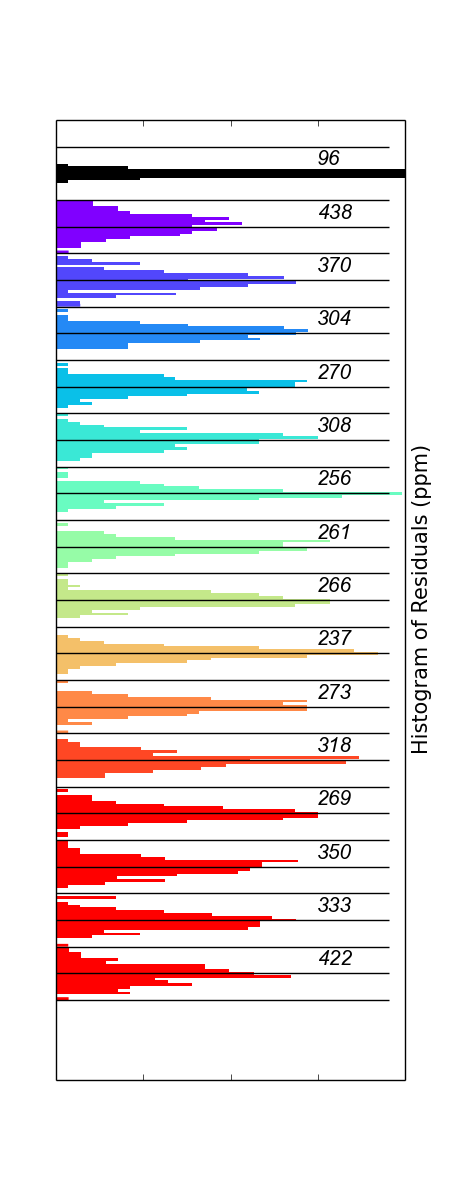}
  \end{minipage}
\vspace{-0.15 in}
\caption{{\it Left:  }  Spectrophotometric transit light curves (colored dots) with systematics removed compared with the best fit transit (solid line).   Light curves are shown as orbital phase (0-1 with 0 as the center of transit) vs. relative flux (vertically shifted for clarity).  Each light curve is labeled with the central wavelength of the spectral bin. {\it Center:  } Residuals to the fit.  {\it Right:  }Normalized histograms of the residuals.  The solid black horizontal lines are spaced at 1000 ppm intervals for scale and the SDNR of each fit is listed in units of ppm.}
\end{figure}

  The systematics were removed from the spectral light curves using the divide white method \citep{2013Deming, 2014Kreidberg,2014Stevenson}.  Removing the best fit transit from the white light curve leaves only the systematics.  We divide each spectral light curve by the systematics.  This assumes that the systematics are wavelength independent. We do note a linear, observation--long, wavelength dependence in the corrected spectral light curves.  We use a first order polynomial to account for the wavelength dependent systematics combined with the transit model when fitting the spectral light curves.  We fix the ephemerides to the white light curve solutions and use fixed, wavelength dependent, non-linear limb darkening coefficients derived in the same way as described in Section \ref{sec: wlc}. The transit depth and normalization factors are left as the only free parameters.  We fit both scan directions simultaneously.  Spectral light curves and their fits are shown in Figure \ref{slc_fits}.

\subsection{Transmission Spectrum}\label{sec: specres}

\begin{figure}[!t]
\centering
\includegraphics[trim=0.04in 0.0in 0.0in 0.0in,clip,width=0.48\textwidth]{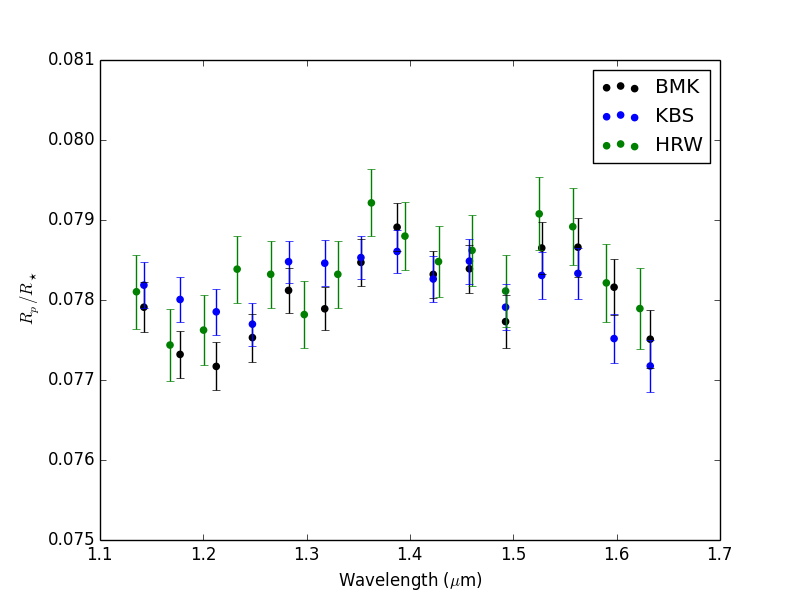}
\caption{Results of spectral extraction from multiple independent analyses show good agreement.  Here we show the best fit value for ${R_p}/{R_{\star}}$ as a function of wavelength with 1$\sigma$ error bars derived from the MCMC posteriors.  Colors correspond to analysis performed by Brian Kilpatrick (BMK), Kevin Stevenson (KBS), and Hannah Wakeford (HRW).}\label{spec_all}
\end{figure}

A transmission spectrum was derived from the transit depth fits of the spectral light curves.  The change in the apparent planetary radius as a function of wavelength can be indicative of absorption features of molecular species in the planetary atmosphere.  As a test for robustness, the spectrum of WASP-63b was extracted using multiple independent analysis pipelines in addition to the method described in detail in the previous subsections (Table \ref{spectab}). Figure \ref{spec_all} shows a comparison of the results from this methodology (BMK) with that of analysis performed using methods described in \cite{2014Stevenson} (KBS) and \cite{2016Wakeford,2017Wakeford} (HRW). The spectra are in good agreement with nearly all points from each of the methods in 1$\sigma$ agreement with each of the other methods.  As a test to determine how small differences in the extracted spectrum would affect retrieval results we analyzed the KBS and HRW spectra with the Pyrat Bay model (Section 3.2.1).  The KBS spectrum has a shallower bump at 1.5 $\mu$m, and thus, the retrieval only detected the water signature significantly (not HCN).  The HRW atmospheric retrieval produced similar results to BMK, but with broader 1$\sigma$ constraints, due to the larger uncertainties of the dataset (compared to BMK or KBS). 

We compared these results against a flat-line model using the Bayesian Information Criterion (BIC, Liddle 2007).  For the atmospheric model of each dataset we searched for the combination of free parameters that minimized BIC.

For KBS, the BIC favors the atmospheric model (BIC=24.59, with three free parameters, $T$, $R_{\rm 0.1 bar}$, and H$\_2$O), against the flat-line model (BIC=30.82, one free parameter).  The posterior odds give the flat-line model a fractional probability of only 0.04 (Raftery 1995).

For HRW, the BIC favors again the atmospheric model (BIC=19.19, with four free parameters, $T$, $R_{\rm 0.1 bar}$, H$\_2$O, and HCN) over
the flat-line model (BIC=22.92).  However, this time there is weaker evidence, with the flat-line model having a fractional probability of 0.13.

Therefore, in all three datasets, there is evidence of spectral features, but with different degrees of confidence.  The BMK spectrum shows strong evidence, KBS spectrum shows moderate evidence, whereas HRW shows only weak evidence for water absorption.  The HRW spectrum is based upon the marginalization technique described in Wakeford 2016 and represents the most conservative estimate of uncertainty.  As such, it serves as a sort of lower bound on the detection threshold whereas the BMK and KBS methods assume we are justified in our choice of systematic model.  The arguments for this choice of model are well extablished and summarized in Section \ref{sec: wlc} and so we will choose the BMK spectral extraction to continue on with retrieval analysis.

\section{Results}\label{sec: res}

The results produced by the methodology described in in detail in Section \ref{sec: OandA}  (presented in Table \ref{spectab} as BMK) were distributed
to the members of the transiting exoplanet community who were involved
with the preparation of \cite{Commtargs} and/or \HST program GO 14642.
Each was given an opportunity to provide an independent analysis of
the results.  Wide community involvement resulted in a number of
contributions in the form of forward model comparisons and retrievals.
Here we present the methods and findings from each interpretation of
the BMK spectrum. All models in this section adopt the system
  parameters from \citet{2012Hellier}.

\subsection{Forward Models}
\label{sec:forward}

\subsubsection{Burrows}

We apply the transit models from
\cite{HoweBurrows2012apjTransitSpectra}, which adopt chemical
equilibrium abundances for molecular species from
\citet{BurrowsSharp1999apjBDandEGP} and opacities from
\citet{SharpBurrows2007apjsBDandEGP}.  The atmospheric models consider
an isothermal temperature profile and gray haze opacity with cross
sections of 0.001--0.005 cm$^{2}$g$^{-1}$ from $10^{-6}$ bar to 1 bar.

By exploring a range of temperatures, haze opacities, metallicities,
and non-equilibrium CO/CH$_{\rm 4}$ abundances, the best-fitting
solutions pointed to solar-abundance atmospheres at a temperature of
1000~K, with a haze/cloud muting the water absorption feature at
1.4 $\mu$m (Figure \ref{fig:retrieval}, top panel).  There is no indication
of significant CH$_{\rm 4}$.  There is a slight degeneracy between the
cloud thickness and temperature, but it is clear that the atmosphere
is cloudy.  These models do not show a significant metallicity
dependence.  Finally, a high CO abundance excess ($\sim$100 times
solar) can help to fit the data at 1.5 $\mu$m, but it does not seem
realistic.

\subsubsection{Heng}

As a complementary approach to the full retrieval calculations, we fit
the data with a 3-parameter analytical model
\citep{HengKitzmann2017isothermalIsobaric}.  In that study, it was
demonstrated that this isothermal, isobaric model matched full
numerical calculations at the $\sim 0.1\%$ level over the WFC3
wavelength range.  The model has three parameters: temperature, water
abundance, and a constant cloud opacity.  The constant cloud opacity
assumes that the cloud particles are large over the wavelength range
probed by WFC3 (i.e., micron-sized or larger radius).  Water opacities
are computed using the HELIOS-K opacity calculator
\citep{GrimmHeng2015apjHELIOSK} and the HITEMP spectroscopic database.
This procedure confronts the data with a simple model, which has a
minimal number of parameters, to serve as a plausibility check.

Following the approach of \citet{KreidbergEtal2015apjWASP12bWater}, we equate
the reference transit radius to the white-light radius and set the
reference pressure to 10 bar.  We assume a hydrogen-dominated
atmosphere and set the mean molecular weight to 2.4.
The top panel of Figure \ref{fig:retrieval} shows the best-fit model
to the WFC3 WASP-63b data.  Our general conclusion mirrors that of the
retrieval calculations: water is present in the atmosphere of
WASP-63b, but its presence is muted by a continuum, which in this case
is attributed to a constant cloud opacity.  The values of our fitting
parameters span a temperature range from 500 to 1000~K, a water
mixing ratio from $\sim 10^{-8}$ to $10^{-7}$, and a cloud opacity
$\sim 10^{-8}$ to $10^{-7}$ cm$^2$\;g$^{-1}$.

\subsubsection{Morley}

\begin{figure}[!htb]
\centering
\includegraphics[width=\linewidth, clip]{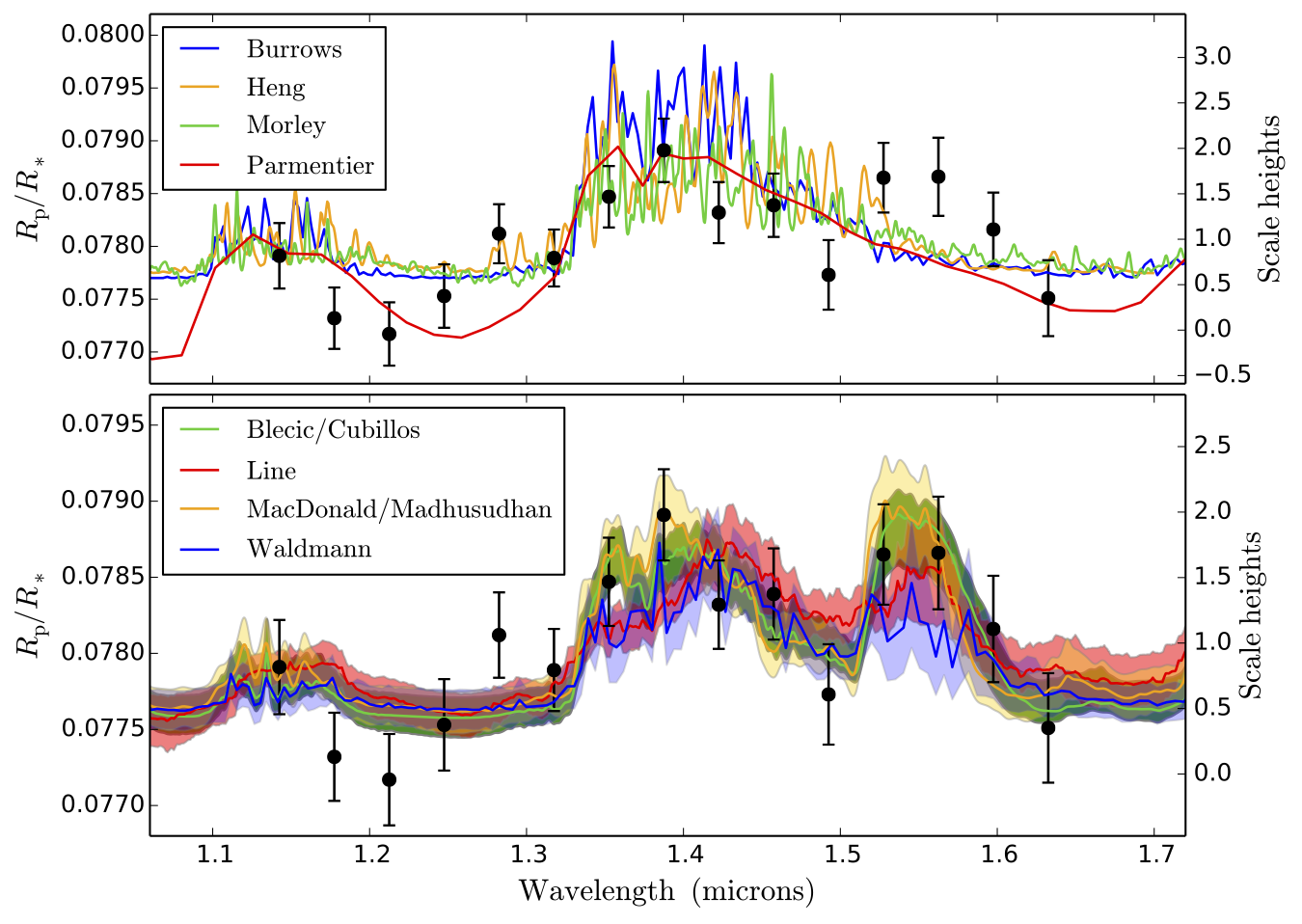}
\caption{
{\it Top:} WASP-63b spectrum and forward models. The black dots with error bars denote the observed best-fit radius ratio and $1\sigma$ uncertainties. The labeled solid curves show representative forward model fits to the data described in section \ref{sec:forward}.
{\it Bottom:} WASP-63b spectrum and retrieval models including HCN absorption. The black dots with error bars denote the observed best-fit radius ratio and $1\sigma$ uncertainties. The labeled solid curves denote the best-fitting models for the consistent retrieval run described in section \ref{sec:discussion}. The vertical shaded areas around each model denote the span of the 1$\sigma$ confidence region of the posterior distribution of sampled models. }
\label{fig:retrieval}
\end{figure}

In order to determine the clouds that are predicted to form in the
atmosphere of WASP-63b and their effect on the planet's transmission
spectrum, we ran self-consistent models including the effects of cloud
condensation.  These models solve for the temperature structure of the
atmosphere in radiative-convective and chemical equilibrium and are
more extensively described in \citet{MckayEtal1989icarTitan,
  MarleyEtal1996sciAtmosphereGliese229B,
  BurrowsEtal1997apjEGPandBrownDwarfs, MarleyEtal1999apjTheorySpectra,
  MarleyEtal2002apjClouds, Fortney2005mnrasTransmissionClouds,
  SaumonMarley2008apjLTdwarfs, FortneyEtal2008apjGPspectra,
  MorleyEtal2015apjFlatModelSpectra}. Our opacity database for gases
is described in \citet{FreedmanEtal2008apjsOpacities,
  FreedmanEtal2014apjsOpacities}, and we calculate the effect of cloud
opacity using Mie theory, assuming spherical particles.  We include
iron and silicate clouds and vary the cloud sedimentation efficiency
$f_{\rm sed}$ from 0.1 to 1, and find that these clouds do indeed form
at high altitudes and damp the size of the signal for low
sedimentation efficiencies (i.e. lofted clouds of small particles).
Figure \ref{fig:retrieval} top panel shows a representative transmission model for
WASP-63b.

\subsubsection{Parmentier}
In order to understand how the three-dimensional structure of the planet might affect our interpretation of the planet's transmission spectrum, we model WASP-63b with the three-dimensional global circulation model SPARC/MITgcm described in~\citet{Showman2009}. Our model solves for the three-dimensional temperature structure of the atmosphere assuming a cloud-free, solar-composition atmosphere. We then use the temperature map to predict the position of the clouds at the limb of the planet by comparing the partial pressure and the saturation pressure of the cloud gaseous constituents as described in~\citet{Parmentier2016}. The cloud top level and size of the cloud particles are free parameters representing vertical mixing  and microphysics respectively. We then compute the transmission spectrum of the whole atmosphere by combining the transmission spectrum obtained with the temperature and cloud profile at each latitude around the limb \citep{parm2018}. 

 Our global circulation model predicts a temperature difference of $400\rm K$ between the east and west limb at the 10 mbar level. As a consequence some cloud species are predicted to be condensed all over the limb of the planet whereas others should condense only on the morning limb and be evaporated on the other one, leading to a partially cloudy atmosphere~\citep{LineParmentier2016apjClouds}. We computed models assuming the presence of enstatite clouds ($\rm MgSiO_3$) or manganese sulfide clouds (MnS) corresponding to a fully cloudy and partially cloud case respectively. Our best fit spectrum with enstatite clouds is very similar to the Morley model of Fig.4. It has a cloud top pressure of 1 mbar and no constraints on the particle size. Our best fit model with MnS clouds is the Parmentier model of Fig. 4. It has a limb that is $\approx 60\%$ cloudy, resulting in a qualitatively different spectrum than the other, one-dimensional models shown here. For this model the cloud top pressure is $\approx0.1$ mbar and the particle size is $\approx 1\mu$m. We conclude that the atmosphere of WASP-63b is unlikely to be clear with a solar-composition abundance of water. Both fully cloudy and partially cloudy atmospheres can exist, depending on the cloud composition. A higher signal to noise spectrum should be able to disentangle between the two scenarios.

\subsection{Retrievals}
\label{sec:retrieval}

Four groups provided atmospheric retrieval analyses for WASP-63b.  The following subsections describe the retrieval procedure and the
individual exploration from each group.  Figures relative to the individual retrievals are included in the Appendix.

\subsubsection{Blecic \& Cubillos}

To model the atmosphere and spectrum of WASP-63b we use the Python
Radiative Transfer in a Bayesian framework (Pyrat Bay)
package\footnote{\href{http://pcubillos.github.io/pyratbay}
  {http://pcubillos.github.io/pyratbay}} (Cubillos et al. 2018, in
prep.; Blecic et al. 2018, in prep.).  Pyrat Bay is an open-source
reproducible code, based on the Bayesian Atmospheric Radiative
Transfer package \citep{Blecic2016phdThesis, Cubillos2016phdThesis}.
The code provides a line-by-line radiative-transfer and a
thermochemical-equilibrium abundances \citep[TEA,
][]{BlecicEtal2016apsjTEA} module, which can be use in forward or
retrieval mode, via a Differential-evolution MCMC sampler
\citep{CubillosEtal2017apjRednoise}.
The atmospheric model consist of a 1D set of concentric shell layers,
in hydrostatic equilibrium, spanning from $10^{-8}$ to $100$ bar.
For the temperature profile we consider either the three-channel
Eddington approximation parameterization
\citep[TCEA,][]{LineEtal2013apjRetrievalI} or an isothermal profile.

The Pyrat Bay code considers molecular opacities for H$_{\rm 2}$O
from HITEMP \citep{RothmanEtal2010jqsrtHITEMP}, NH$_{\rm 3}$ and
CH$_{\rm 4}$ from HITRAN \citep{Rothman2013JqsrtHITRAN}, and HCN
from Exomol \citep{Barber2014mnrasExomolHCN}.
We compressed these line-by-line data files using the
  \textsc{repack} package \citep{Cubillos2017apjRepack} to
  extract only the strong lines that dominate the spectrum, speeding
  up the radiative-transfer calculation.
Additionally, Pyrat Bay considers
collision induced
absorption (CIA) from H$_{\rm 2}$-H$_{\rm 2}$
\citep{BorysowEtal2001jqsrtH2H2highT, Borysow2002jqsrtH2H2lowT} and
H$_{\rm 2}$-He \citep{BorysowEtal1988apjH2HeRT,
  BorysowEtal1989apjH2HeRVRT, BorysowFrommhold1989apjH2HeOvertones};
and H$_{\rm 2}$ Rayleigh scattering
\citep{LecavelierDesEtangsEtal2008aaRayleighHD189}.  We also consider
two cloud parameterizations, a simple gray-cloud opacity with constant
cross section (cm$^{-2}$ molec$^{-1}$) below $10^{-5}$ bars, and a
thermal-stability cloud model based on the approach described in
\citet{AckermanMarley2001apjCondensates} and
\citet{Benneke2015arxivCOretrieval}, with additional flexibility
(Blecic et al. 2018, in prep).  We compute the opacity for either Fe
or MgSiO$_3$ condensates using Mie-scattering theory
\citep{ToonAckerman1981aoScattering}.

The retrieval parameterization includes free scale factors for the
  mixing ratios of H$_2$O, NH$_{\rm 3}$, CH$_4$, and HCN
  (vertical-constant values) and the mean molecular mass of the
  atmosphere; either the isothermal temperature or the $\kappa$,
  $\gamma$, and $\beta$ parameters of the TCEA model \citep[see
][]{LineEtal2013apjRetrievalI}; the gray cloud cross section or
  the Mie-scattering cloud profile shape, condensate mole fraction,
  particle-size distribution, and gas number fraction just below the
  cloud deck.; and the planetary radius at 0.1 bar.

We explored several cases, obtaining qualitatively good fits in all
gray-cloud, complex-cloud, and clear-atmosphere cases.  As expected
for transmission spectroscopy, the retrieval returned largely
unconstrained parameters for the TCEA temperature model, suggesting
that the data does not justify more complex models than an isothermal
profile.  In all cases the MCMC favors lower temperatures ($T <
1000$~K) than equilibrium temperature (1500~K) at the pressures probed
by the observations.  We constrain the water abundance, ranging from
solar to $\sim$0.1 times solar values.  The observed water absorption
feature is muted relative to a clear atmosphere with solar abundances.
This is caused by a sub-solar water abundance, an absorbing cloud
opacity, or a high mean molecular mass, which reflects in a strong
correlation between the water abundance and the cloud cross-section.
When we compare retrievals with the gray and complex cloud model, we
find similar best-fitting spectra between the two cases.  The
complex-cloud retrieval does not constrain any of the cloud parameters
when we set all four species abundances free.  In the case when we set
the water abundance as the only abundance free parameter, we find a
somewhat better condensate-fraction constraint.  Since the cloud
opacity dominates only a limited region of the observed spectrum
($\sim$1.2--1.3 $\mu$m), we conclude that there is no need for a more
complex cloud model for this study.
The Reproducible Research Compendium for the Pyrat
    Bay models is available at
\href{https://github.com/pcubillos/KilpatrickEtal2018\_WASP63b}
     {https://github.com/pcubillos/KilpatrickEtal2018\_WASP63b}.

\subsubsection{Line}

We use the CHIMERA transmission model
\citep{LineEtal2013apjHATP12b, SwainEtal2014apjHD189733b,
  2014Kreidberg, KreidbergEtal2015apjWASP12bWater,
  GreeneEtal2016apjAtmospheresJWST, 2016Line}.  For
transit geometry, the code solves the radiative-transfer equation for
parallel rays across the terminator of the planet
\citep{Brown2001apjTransmissionSpectra, TinettiEtal2012rsptaWater}.
The code integrates
atmospheric opacities from either correlated-K or sampled ``line-by-line"
absorption cross sections.
To explore the parameter space, the transmission model is coupled with
the PyMultiNest \citep{BuchnerEtal2014aaPyMultiNest} multimodal
nested-sampling algorithm \citep{FerozHobson2008mnrasNestedSampling}.
The atmosphere is in hydrostatic equilibrium with height-dependent
gravity, temperature, and molecular weight.  The temperature profile
comes from either the radiative-equilibrium model from
\citet{Guillot2010aaRadEquilibrium} or an isothermal profile,
  spanning from $10^{-7}$ to 30 bar.
The atmosphere uses 'thermochemically-consistent' molecular abundances
\citep[as defined in][]{KreidbergEtal2015apjWASP12bWater}, computed
using the NASA CEA2 model for given elemental abundances
\citep{Lodders2009ExoClemistry}.

The CHIMERA code considers molecular opacities for H$_2$O, CH$_4$,
  CO, CO2, NH$_3$, Na, K, TiO, VO, C$_2$H$_2$, HCN, and FeH
\citep{FreedmanEtal2014apjsOpacities}; CIA from H$_{\rm
    2}$-H$_{\rm 2}$ and H$_{\rm 2}$-He
\citet{RichardEtal2012jqsrtCIA}; a Rayleigh power-law haze
\citep{LecavelierDesEtangsEtal2008aaRayleighHD189}; and either an
opaque gray patchy cloud model \citep{LineParmentier2016apjClouds} or
a more complex, Mie-cloud model
\citep{LeeEtal2013apjAtmosphericRetrieval}.

The retrieval parameterization includes the metallicity $\rm
   [M/H]$, the carbon-to-oxygen ratio $\log(\rm C/O)$, and the carbon-
   and nitrogen-species quench pressures
   \citep{KreidbergEtal2015apjWASP12bWater, MorleyEtal2017ajGJ436b} to
   set the elemental abundances; either the isothermal temperature
     or the $\kappa_{\rm IR}$, $\gamma_v$, $T_{\rm irr}$ parameters
     of \citep{Guillot2010aaRadEquilibrium}; the top-pressure
   boundary and a 'patchy terminator' parameter for the gray patchy
   cloud model, or the $Q_0$ and $r$ \citep[see
   ][]{LeeEtal2013apjAtmosphericRetrieval} and profile shape
     (mixing ratio, cloud base pressure, pressure fall off index, Line
     et al. 2017, in prep.) for the Mie-cloud model; and a radius
   scale factor to set the reference altitude at 10 bar.

The 'chemically-consistent' retrieval detects the water spectral
feature at $3.6\sigma$ confidence.
This is consistent with a hard upper limit on C/O near 1.  The water
band is muted relative to solar composition.  The retrieval posterior
shows two 'composition' modes: low metallicity (${\rm [M/H]} \lesssim
1.3$ (20$\times$)) degenerate with a cloud and high metallicity (peak
near $\sim$300$\times$solar).  There is a turn-over degeneracy in
cloud top vs. [M/H] (due to the effect on the mean molecular weight)
resulting in the bi-modal marginalized metallicity distribution.
Clouds can be present, but are not required to fit the spectra as
given by the Bayes factor (0.45) and result in a much lower value for
the low metallicity mode (<0.1$\times$solar), while the high
metallicity mode remains.
The highest of the sampled metallicities (greater than $\sim$50 times
solar) are possibly implausible given mass and radius of WASP-63b.

Further tests found negligible variations when imposing a temperature
prior or no patchy-cloud factor.  A comparison between correlated-K
and line-by-line sampling opacities produced nearly identical results.
Likewise, the more complex Mie-cloud model produced qualitatively
similar main conclusions (with unconstrained cloud particle sizes,
vertical extent, or cloud composition).

\subsubsection{MacDonald \& Madhusudhan}

We use the nested-sampling retrieval algorithm POSEIDON
\citep{MacDonaldMadhusudhan2017poseidon} to analyze the WFC3
observations of WASP-63b.  The code pre-computes line-by-line
  molecular cross sections following the methodology of
\citet{HedgesMadhusudhan2016mnrasBroadening} and
\citet{GandhiMadhusudhan2017genesis}.  To compute detection
significances we conduct nested Bayesian model comparisons.
For simplicity, we model the atmosphere assuming an isothermal
temperature-pressure profile, with 100 layers uniformly spaced in
log-pressure from $10^{-6}$ to $100$ bar, in hydrostatic equilibrium.

The POSEIDON code considers molecular opacities for H$_2$O from HITEMP
\citep{RothmanEtal2010jqsrtHITEMP} and CH$_4$, NH$_{\rm 3}$, and HCN
from Exomol \citep{TennysonEtal2016jqsrtExomol}; CIA from H$_{\rm
    2}$-H$_{\rm 2}$ and H$_{\rm 2}$-He
\citep{RichardEtal2012jqsrtCIA}; Rayleigh scattering
\citep{LecavelierDesEtangsEtal2008aaRayleighHD189}; and a
uniform-in-altitude gray opacity cloud model.

The retrieval parameterization includes free scale factors for
the mixing ratio of H$_2$O, CH$_4$, NH$_{\rm 3}$, and HCN;
the isothermal temperature; the gray-cloud opacity;
and the reference pressure at the transit radius.

The model comparison test marginally prefer the gray-opacity case
($\chi^2_{\rm red}=1.21$) over a cloud-free case ($\chi^2_{\rm
  red}=1.46$), with a Bayes factor of 2.2.  Adopting the gray-opacity
model, we detect H$_2$O at $4.0 \sigma$ (Bayes factor = 601), HCN at
$3.1 \sigma$ (Bayes factor = 27.6), and 'nitrogen chemistry'
(combination of HCN and NH$_{\rm 3}$) at $3.3 \sigma$ (Bayes factor =
53.7).  We do not detect CH$_4$.

\subsubsection{Waldmann}

We retrieved the \HST/WFC3 spectrum of WASP-63b using the Tau-REx
atmospheric retrieval framework \citep{WaldmannEtal2015apjTauRexI,
  WaldmannEtal2015apjTauRexII, Waldmann2016apjRobert}.  Based on the
Tau code transmission forward models by
\citet{HollisEtal2013cophcTAU}, Tau-REx employs Nested Sampling
\citep{FerozHobson2008mnrasNestedSampling} to solve the full Bayesian
argument.  Tau-REx can use high-resolution absorption cross-section or
correlated-k tables as opacity inputs.  Here we used the latter but
find both to yield equivalent results for the wavelength ranges and
sensitivities of the data at hand.  We include pressure-dependent line
broadening where such information is available, taking into account
the J quantum number dependence on pressure broadening coefficients.
In this study, we assume an isothermal atmospheric
temperature-pressure profile, spanning from to $10^{-9}$ to 10
  bar.

The Tau-REx code considers molecular opacities for
  H$_2$O, CH$_4$, NH$_3$, HCN, TiO, and VO from Exomol
\citep{TennysonYurchenko2012mnrasExoMol},CO and CO2 from HITEMP
\citep{RothmanEtal2010jqsrtHITEMP}, and O$_2$, O$_3$, NO$_2$, NO,
  HCOOH, C$_2$H$_6$, and C$_2$H$_2$ from HITRAN
\citep{RothmanEtal2009jqsrtHITRAN, Rothman2013JqsrtHITRAN}; CIA from
H$_2$-H$_2$ \citep{BorysowEtal2001jqsrtH2H2highT,
  Borysow2002jqsrtH2H2lowT} and H$_2$-He \citep{AbelEtal2012ciaH2-He};
Rayleigh scattering
\citep{LecavelierDesEtangsEtal2008aaRayleighHD189}; and clouds using a
hybrid model of gray-cloud opacities and a phenomenological Mie
scattering \citep{LeeEtal2013apjAtmosphericRetrieval}.

We run two types of retrievals, a 'free' retrieval with abundances of
H$_2$O, CH$_4$, CO, CO$_2$, NH$_3$, as well as a chemical-equilibrium
retrieval using an implementation of the ACE model by
\citet{AgundezEtal2012aaChem}, with the C/O ratio and atmospheric
metallicity as free parameters.  The rest of the retrieval parameters
are the isothermal temperature; the top pressure of the gray cloud
model and the particle size, composition and mixing ratio of the
Mie-cloud model; and the planet reference radius at 10 bar.

\begin{figure*}
\centering
\includegraphics[trim=0 0 0 0, clip, height=0.38\linewidth,
clip]{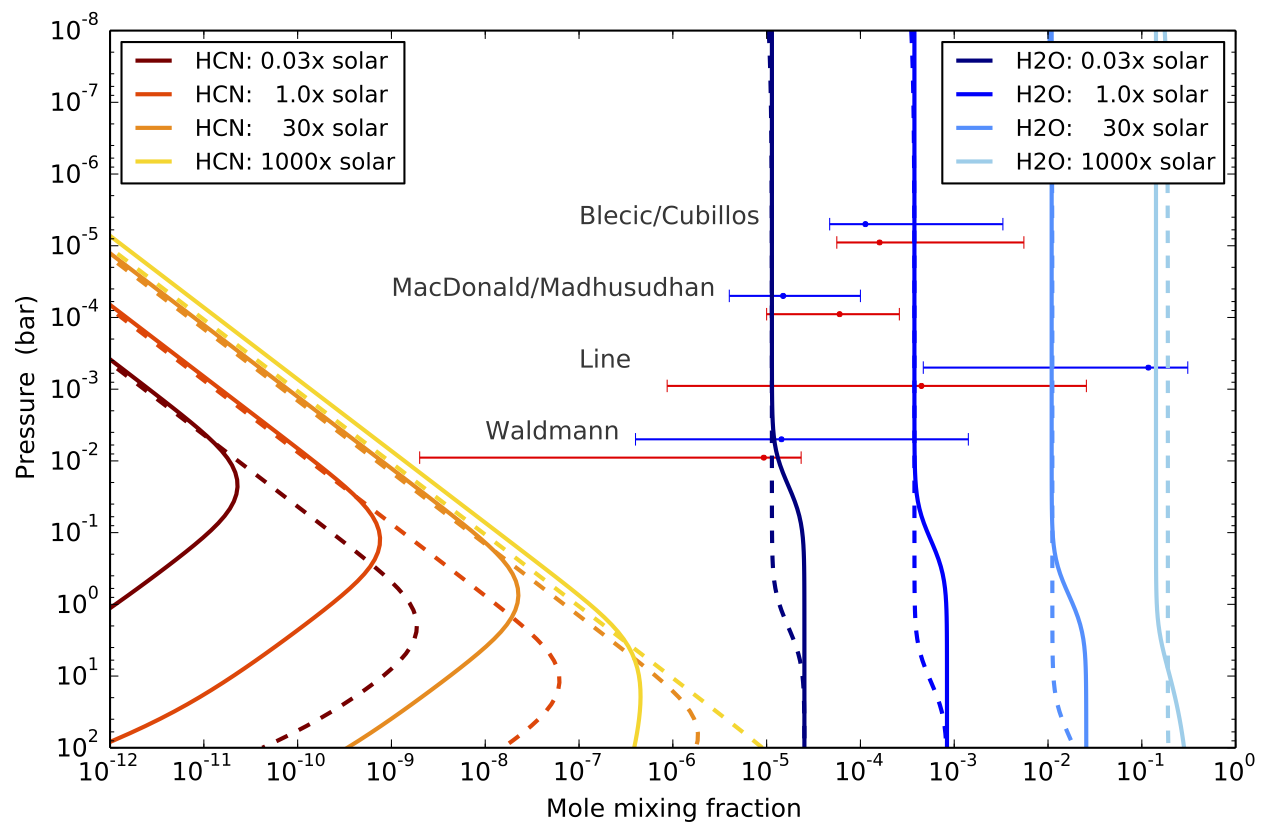}
\includegraphics[trim=0 -10 5 0, clip, height=0.38\linewidth, clip]{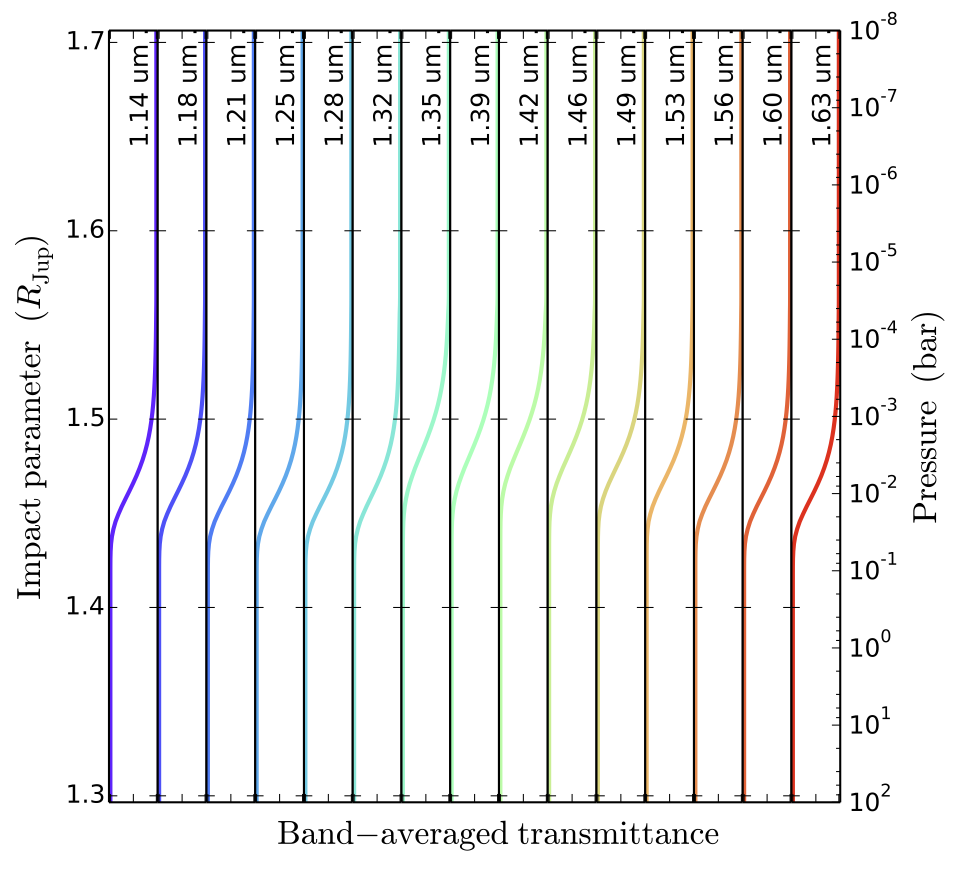}

\caption{{\it Left:} water and HCN abundances for WASP-63b compared
  to thermochemical-equilibrium mole mixing fractions models for an
  isothermal atmosphere at 1000~K (solid lines) and at 1500 K (dashed
  lines) at a range of metallicities (see labels). The horizontal
  error bars show the retrieved 68\% credible-region abundances from
  each group (labels) for water (blue) and HCN (red).  \textbf{Note
    that the retrieved abundances are vertically offset for clarity, see right panel for the probed pressures.}
  All abundances correspond to the pressure levels probed by this
  transmission observation (10$^{-2}$--10$^{-4}$ bar).  All retrieved
  water abundances are consistent with solar or slightly sub-solar
  water abundances, however the retrieved HCN values range several
  order of magnitudes higher than any 1000 K equilibrium value
  (particularly for the solar to sub-solar HCN curves).  The
  retrieved mole mixing fractions of HCN are more plausible if the
  temperature at deeper layers were $\sim$1500 K and vertical
  transport dominated the abundances in the region probed by
  observations. The models suggest it would be possible under these
  conditions to produce mole mixing fractions on the order of
  10$^{-6}$. {\it Right:} Sample transmittance curves as a
    function of impact parameter for each WFC3 pass band (see labeled
    wavelengths) for the Blecic/Cubillos retrieval best fit.  The
    transition from optically thin to optically thick between
    $\sim$1.5 an $\sim$1.4 $R_{\rm Jup}$, respectively, denotes the
    photosphere of the planet, as observed by WFC3. The pressure scale
    on the right side denotes the deepest atmospheric pressure probed
    by each impact parameter.  The transmittance curves from other
    groups show a similar trend.}
\label{fig:abundances}
\end{figure*}

We detect water with a $3.5 \sigma$ significance (Bayes factor = 103)
compared with a family of pure-cloud or  featureless atmosphere models.  We
obtain $\log({\rm H}_2{\rm O}) = -4.84^{+1.04}_{-1.53}$.  We do not find any
evidence of an extended Rayleigh curve due to hazes but found a
gray-cloud model to be sufficient.  We constrain the cloud-top
pressure to $\log(p) = 3.08^{+1.48}_{-0.93}$ Pa.
The chemically consistent retrieval yielded two results.  The first result yields a high metallicity atmosphere at 370 times solar.  The
second result yields a low metallicity atmosphere at 0.24 times solar.
Both solutions feature comparable log-evidences and result in upper bounds of C/O at 0.49.  A ratio of ${\rm C/O} < 0.7$ is expected as only
water is retrieved in this data set and is therefore consistent with the 'free' retrieval approach above.  The atmospheric metallicity is
poorly constrained due to the presence of clouds which has the effect of muting the water feature and biasing the chemical-consistent model
to either compensate with unrealistically high mean molecular weight atmospheres or unrealistically low trace gas abundances.

\section{Discussion}
\label{sec:discussion}

The individual atmospheric analyses of WASP-63b agree that there is a robust water detection (3.5--4.0 $\sigma$), but with a muted absorption feature when compared to a clear solar-composition atmospheric model.  It is unclear if the reason for the muting of the feature is the result of sub-solar water abundance, absorbing cloud opacity, or a high mean molecular mass.  Thermochemically-consistent retrievals show a multimodal solution due to degeneracies between cloud opacity and composition (Figures: \ref{fig: lineapp} and \ref{fig: waldapp}).  Retrievals with simple temperature (isothermal) and cloud (gray opacity) models both produced fits consistent with retrievals with more complex models and thus the data does not warrant the incorporation of more complex models nor does it allow further constraints on cloud properties.

The high transit-depth values between 1.5 and 1.6 $\mu$m motivate the inclusion of HCN and the exploration of disequilibrium chemistry.  Each retrieval team performed an additional retrieval exercise with a common set of assumptions to further explore the the inclusion of HCN as a means to fit the `bump' in the spectrum between 1.5 and 1.6 $\mu$m.  We implemented an isothermal temperature model, a gray-opacity cloud model, a free pressure--radius reference point, and opacities from H$_2$-Rayleigh,
H$_2$-H$_2$ and H$_2$-He CIA, and H$_2$O, CH$_4$, HCN, and NH$_3$.  We adopted molecular abundances either from thermochemically-consistent calculations or from free abundances (constant vertical profiles), with the exception of the HCN abundance, which is always a free fitting variable (constant vertical profile). 

We begin with this common set of assumptions and then compare retrieval results from teams with differing retrieval frameworks. Notable differences include:  the statistical sampler from Blecic/Cubillos (MCMC) differ from the rest (Nested sampling), the molecular opacity handling from Waldmann and Line (correlated-K) differ from the others (cross-section sampling), and the chemistry from Line (thermochemically-consistent) differs from the rest (free constant vertical profiles).  Figure \ref{fig:retrieval} shows the retrieved spectrum from the run using the common assumptions.  All four retrievals produced consistent spectral fits, seen in the intersecting 68\% confidence regions around the best-fitting models.  In terms of the atmospheric characterization, these retrievals confirm the previously found water detection.  

The Bayesian hypothesis testing favors the fit with HCN, improving the fit at 1.5--1.6 $\mu$m.  However, the detection significance is low and inconsistent, $3.1\sigma$ (Blecic/Cubillos), $2.1\sigma$ (Line), $3.1 \sigma$ (MacDonald/Madhusudhan), and $1.9 \sigma$ (Waldmann).  Therefore, for the currently available data, the inclusion of HCN is not statistically justified within this model parameterization.  Furthermore, to reproduce the observed values requires the HCN mole fraction be  $\gtrsim 10^{-5}$; much higher than thermochemical-equilibrium values at the pressures probed by the WFC3 observations (Figure \ref{fig:abundances}).
To produce such high HCN abundances, one would need to invoke disequilibrium-chemistry processes by either quenching or photochemistry.  Quenching can occur when higher temperatures at deep  layers, below the levels probed by this observation, enhance the HCN abundance without needing the high metallicities from Figure \ref{fig:abundances}.  If vertical mixing dominates the mid-altitude abundances of the WASP-63b atmosphere (expected at the retrieved temperatures of $\sim$1000~K), HCN could be effectively quenched, maintaining the high abundances from the deep layers throughout the probed region.  Similar deviations from equilibrium chemistry have been modeled for other Jupiter-like exoplanets \citep{MosesEtal2011apjDisequilibrium,
  VenotEtal2012aaHotJupiterChemistry}.  None of the retrievals constrain any of the other molecular abundances that could provide additional evidence for quenching (e.g. CO, CH$_{4}$, NH$_{3}$).  However, photochemistry could play a role in removing these other molecules from the atmosphere while enhancing the mole fraction of HCN at pressures less than a millibar.  \cite{MosesEtal2011apjDisequilibrium,2013Moses}  show that ammonia and methane can be photochemically converted to HCN at the pressure levels probed by near--IR transmission spectroscopy thus driving the retrieved abundances much higher than equilibrium values.
Future observations with extended wavelength ranges and higher sensitivity, such as \JWST, can help to definitively confirm or rule out the detection of HCN, and other atmospheric species, thus constraining the presence of disequilibrium chemistry.

\section{Conclusions}
We present the observations of one transit of the hot Jupiter WASP-63b.  Observations were conducted in  the  near-infrared  using  \HST  WFC3  G141.   This study was done as a preliminary evaluation of the suitability of  WASP-63b as one of the  community  targets  for \JWST ERS science.  We have detected a muted water absorption feature at $\sim$ 1.4 $\mu$m confirming WASP-63b as a potential target for ERS science. The potential presence of an absorption feature at 1.55$\mu$m is not evidence enough to make strong conclusions about the presence of other molecules in the atmosphere, however, further observations by \JWST would be able to identify additional spectral features that would allow us to further constrain the atmospheric composition. The observational window for observing WASP-63b with \JWST is from September 23 -- April 5.  \JWST is currently scheduled to launch in October 2018 and ERS observations would commence in April 2019.  Assuming the mission remains on schedule, WASP-63b would not be observable until several months after the ERS program window.  However, if there are any delays to launch or the start of ERS observations, WASP-63b would be a prime candidate for study with multiple instruments and modes.\\

\section{Acknowledgments}
This work is based on observations made with the NASA/ESA Hubble Space Telescope, obtained from the Data Archive at the Space Telescope Science Institute, which is operated by the Association of Universities for Research in Astronomy, Inc., under NASA contract NAS 5-26555. These observations are associated with program GO-14642. BMK acknowledges funding by HST-GO-14642.047 provided by the Space Telescope Science Institute.  H.R. Wakeford acknowledges support from the NASA Postdoctoral Program, administered by USRA through a contract with NASA.  Jasmina Blecic is supported by NASA trough the NASA ROSES-2016/Exoplanets Research Program, grant NNX17AC03G.  \par The authors acknowledge the contributions and support from members of the transiting exoplanet community who have contributed to and/or supported Program GO-14642 including: E. Agol, D. Angerhausen, T. Barman, J. Barstow, N. M. Batalha, S. Birkman, D. Charbonneau, N. Cowan, N. Crouzet, S. Curry, J. M. Desert, D. Dragomir, J. Fortney, A. Garcia Munoz, N. Gibson, J. Gizis, T. Greene, J. Harrington, T. Kataria, E. Kempton, H. Knutson, L. Kreidberg, M. Lopez-Morales, M. Rocchetto, E. Schlawin, E. Shkolnik, A. Shporer, D. Sing, K. Todorov, and J. de Wit. 

\bibliography{w63bib}

\newpage
\cleardoublepage

\appendix\label{app}
Here we present the retrieval results as described in Section \ref{sec: res}.  We present the pairs plots and fit to the observations in each case.   In the case of the Line and Waldmann results we show both the thermochemically-consistent run along with the free retrieval for comparison.  \\

\section{Blecic \& Cubillos}
\begin{figure}[!h]\label{fig: pcapp}
\centering
\includegraphics[width=0.60\linewidth]{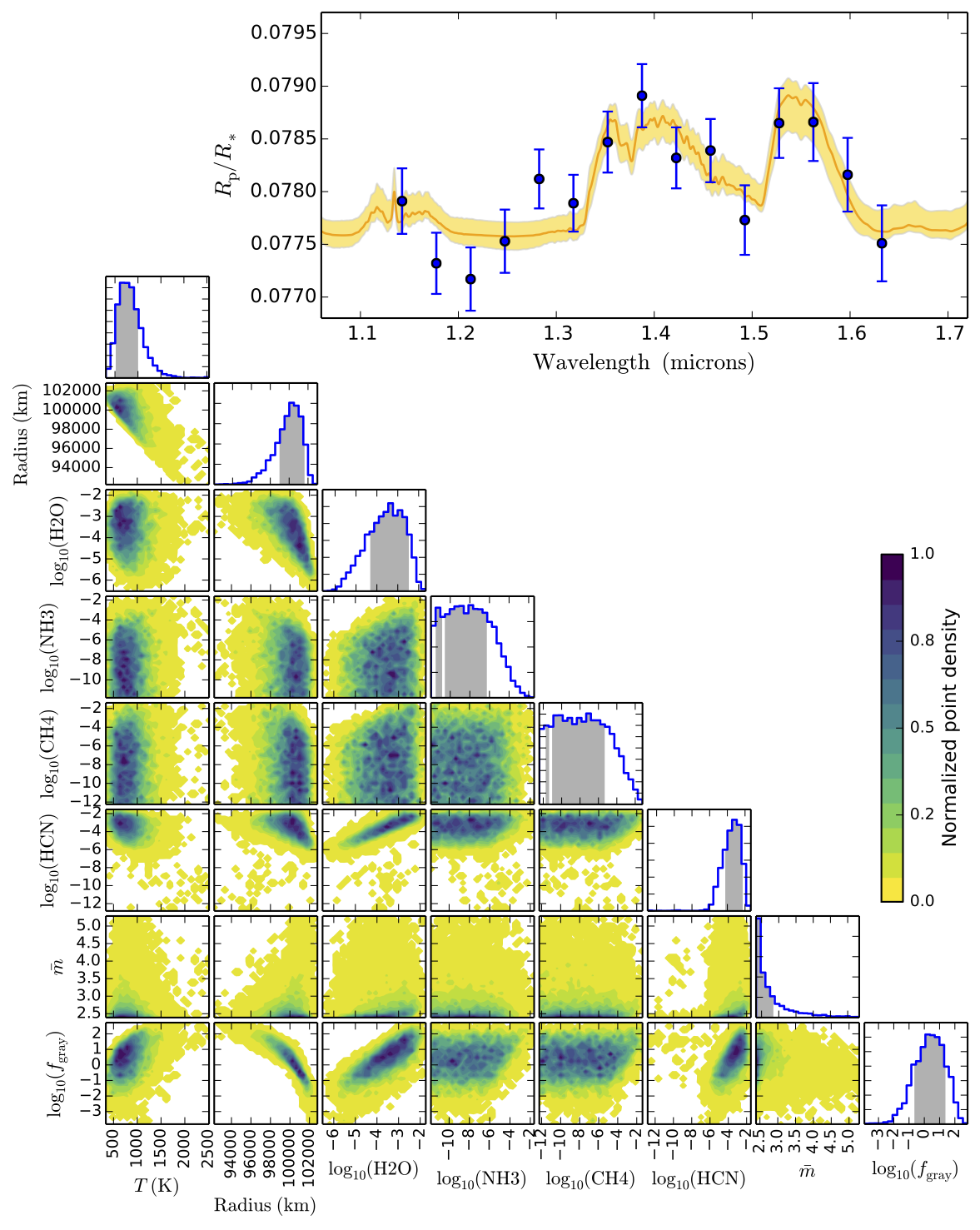}
\caption{Posteriors and fit from retrieval parameterized by free abundances of H$_{\rm 2}$O, NH$_{\rm 3}$, CH$_{\rm 4}$, and HCN along with the mean molecular mass of the atmosphere.  The atmosphere is assumed to be isothermal ($T$ as a free parameter) with a grey cloud (opacity as free parameter).}
\end{figure}

\clearpage

\section{Line}
\begin{figure}[!h]\label{fig: lineapp}
\centering
\includegraphics[ width=0.9\linewidth]{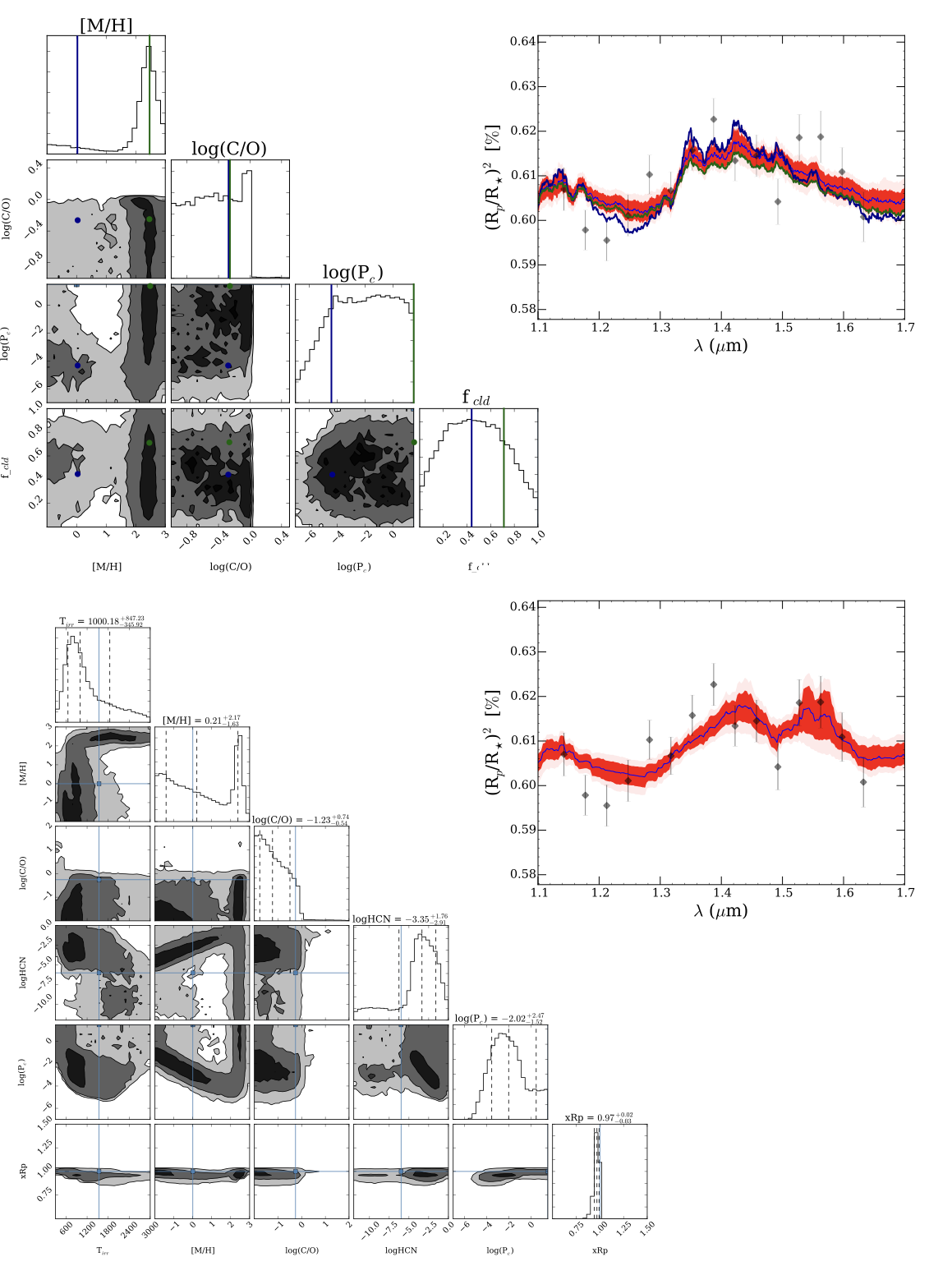}
\caption{{\it Top:  }Posteriors and fit to spectrum of thermo-chemically consistent retrieval. The elemental abundances are parameterized by the metallicity $\rm
[M/H]$, the carbon-to-oxygen ratio $\log(\rm C/O)$, and the carbon- and
nitrogen-species quench pressures. {\it Bottom: }Posteriors and fit to spectrum of assuming a thermo-chemically consistent atmosphere with the addition of HCN as a free parameter.}

\end{figure}
\clearpage

\section{MacDonald \& Madhusudhan}
\begin{figure}[!h]\label{fig: mmapp}
\centering
\includegraphics[height=0.3\textheight, width=0.7\linewidth]{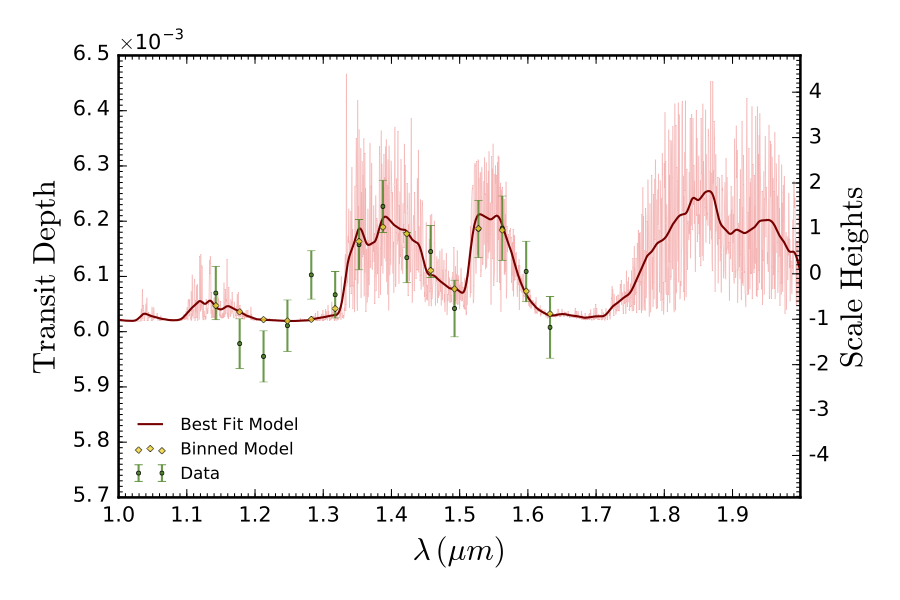}
\includegraphics[height=0.5\textheight, width=0.7\linewidth]{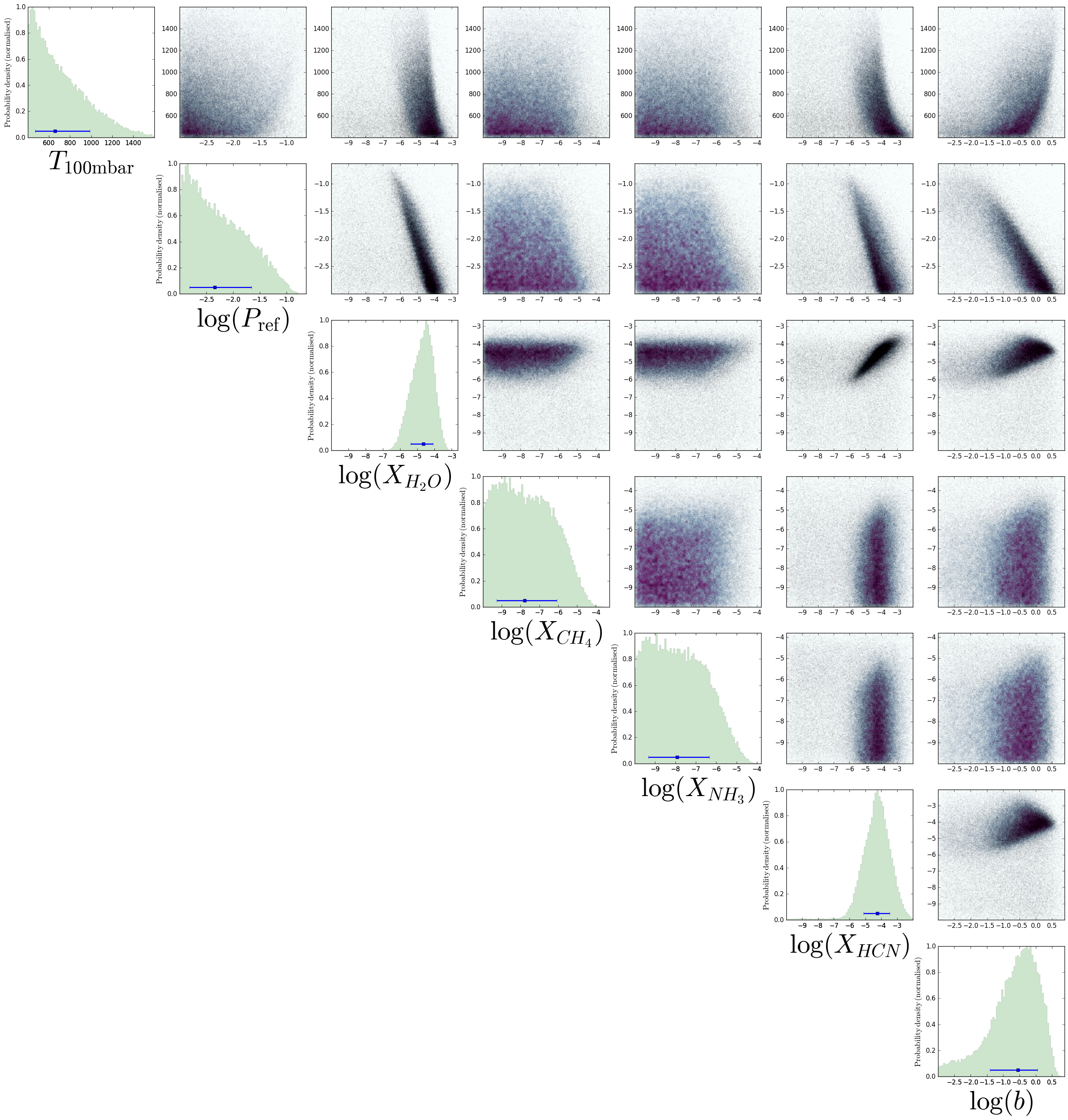}\\
\caption{{\it Top:  } Fit to spectrum of free retrieval. {\it Bottom:  }Posteriors of free retrieval assuming an isothermal temperature-pressure profile, including molecular
opacities due to H$_2$O, CH$_4$, NH$_{\rm 3}$, and HCN, and clouds as a uniform-in-altitude gray opacity.}

\end{figure}
\clearpage
\newpage

\centering
\section{Waldmann}

\begin{figure}[!h]\label{fig: waldapp}
\includegraphics[trim=0.0in 0.0in 0.0in 0.0in,clip, width=0.55\textwidth, angle = 270]{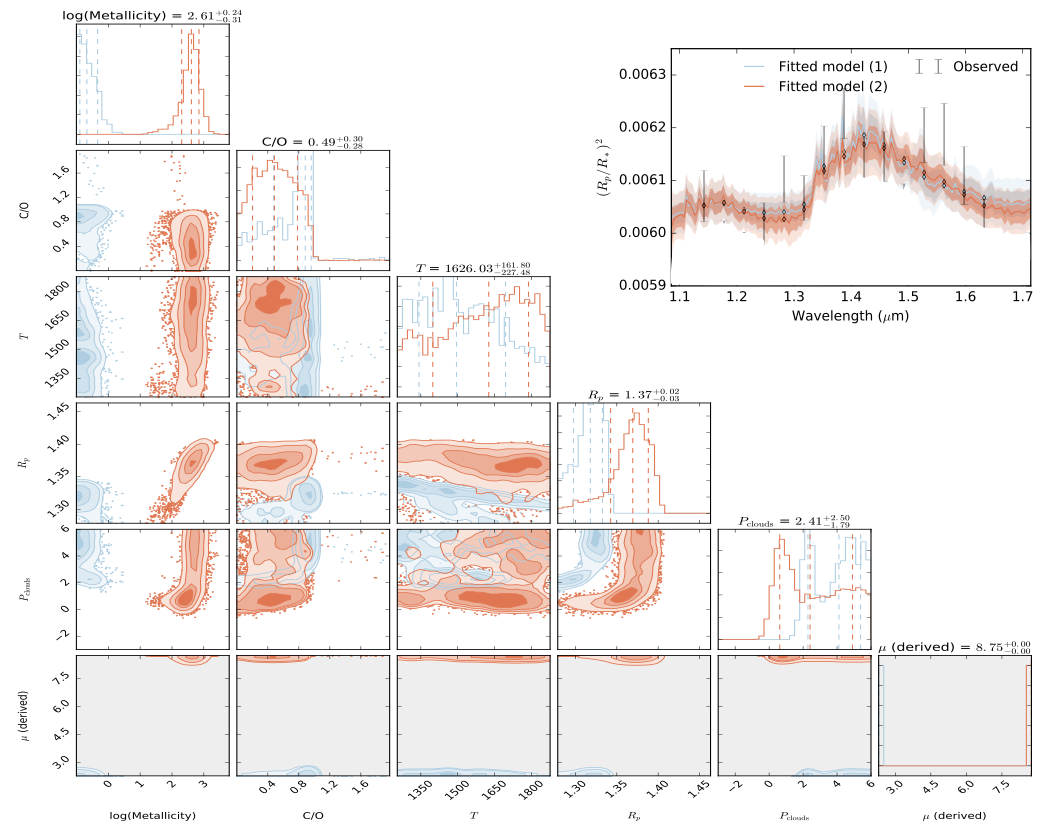}\\

\includegraphics[trim=0.0in 0.0in 0.0in 0.1in,clip,width=0.55\linewidth, angle = 270]{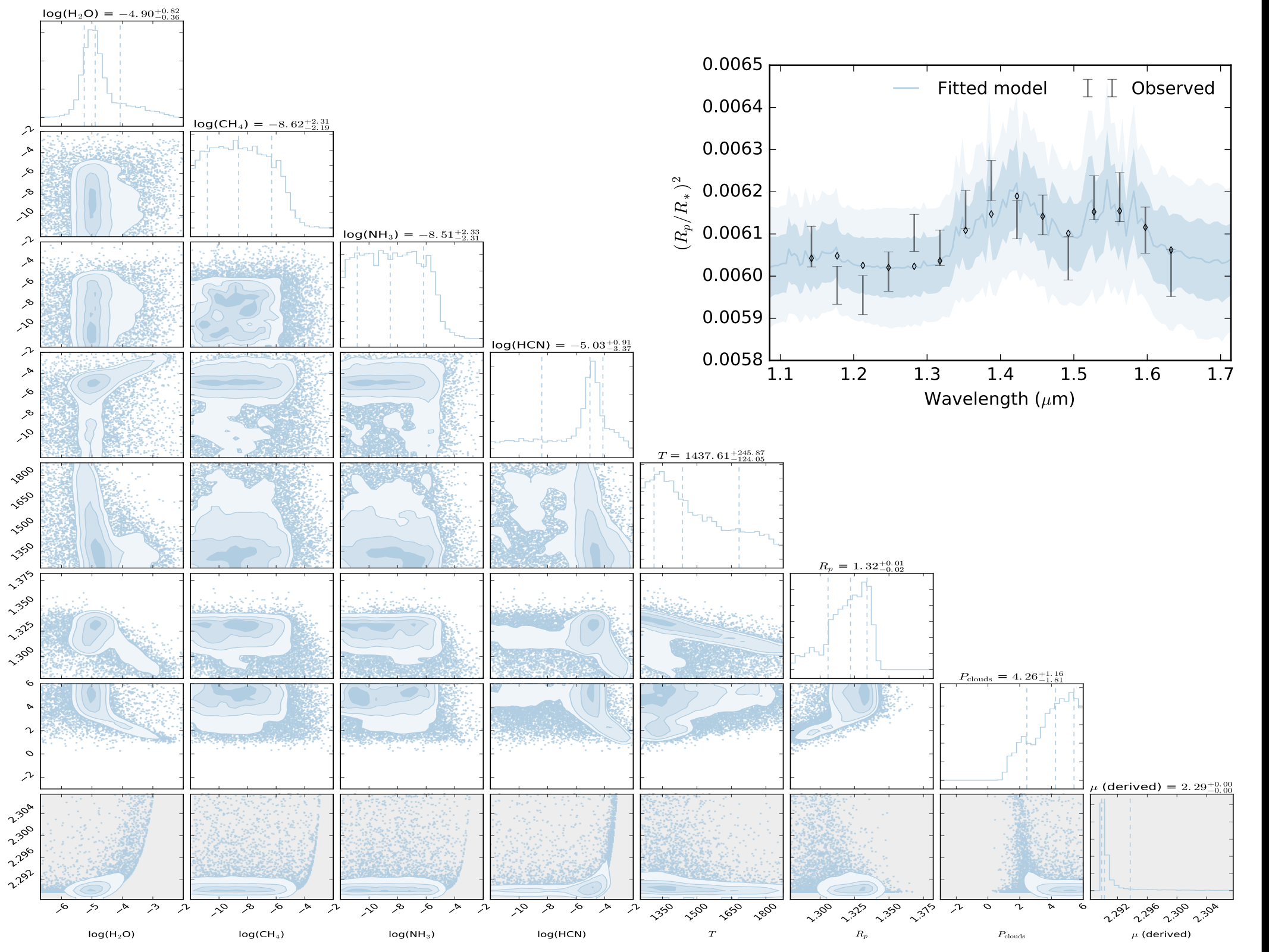}

\caption{{\it Top:  }Posteriors and fit to spectrum of chemically consistent retrieval.  The free parameters are the C/O ratio, atmospheric metallicity, planet radius, temperature and
cloud-top pressure.  {\it Bottom: }Posteriors and fit to spectrum of free retrieval with planet radius,
temperature, cloud-top pressure and abundances of H$_2$O, CH$_4$, CO,
CO$_2$, NH$_3$, and HCN as free parameters.}
\end{figure}

\end{document}